\documentclass{IEEEtran}
\usepackage{latexsym}
\usepackage[dvips]{graphicx}
\usepackage{booktabs}
\usepackage{amsmath}
\usepackage[dvips]{psfrag}
\usepackage{psfrag}
\begin{document}

\title{Physical Layer Network Coding: A Cautionary Story with Interference and Spatial Reservation}

\author{\IEEEauthorblockN{Hironori Fukui\IEEEauthorrefmark{1}, Hiroyuki Yomo\IEEEauthorrefmark{1}\IEEEauthorrefmark{2}, and Petar Popovski\IEEEauthorrefmark{2}}

\IEEEauthorblockA{\IEEEauthorrefmark{1}Graduate School of Science and Engineering, Kansai University, Japan}

\IEEEauthorblockA{\IEEEauthorrefmark{2}Department of Electronic Systems, Aalborg University, Denmark}}

\maketitle
\thispagestyle{empty}
\begin{abstract}
Physical layer network coding (PLNC) has the potential to improve throughput of multi-hop networks.
However, most of the works are focused on the simple, three-node model with two-way relaying, not taking into account the fact that there can be other neighboring nodes that can cause/receive interference. The way to deal with this problem in distributed wireless networks is usage of MAC-layer mechanisms that make a spatial reservation of the shared wireless medium, similar to the well-known RTS/CTS in IEEE 802.11 wireless networks. In this paper, we investigate two-way relaying in presence of interfering nodes and usage of spatial reservation mechanisms. Specifically, we introduce a \emph{reserved area} in order to protect the nodes involved in two-way relaying from the interference caused by neighboring nodes. We analytically derive the end-to-end rate achieved by PLNC considering the impact of interference and reserved area. A relevant performance measure is data rate per unit area, in order to reflect the fact that any spatial reservation blocks another data exchange in the reserved area. The numerical results carry a cautionary message that the gains brought by PLNC over one-way relaying may be vanishing when the two-way relaying is considered in a broader context of a larger wireless network. 
\end{abstract}

\section{Introduction}
Physical layer network coding (PLNC) is a promising approach to improve throughput in multi-hop networks and has been 
extensively studied in literature\cite{PLNC_pp_hy}-\cite{Zhang}.
With PLNC, two nodes simultaneously transmit packets to a relay.
The relay amplifies the received signal\cite{PLNC_pp_hy}\cite{analog} or applies symbol mapping\cite{PopovskiICC}\cite{Zhang} based on the received, combined signal, and then broadcasts the processed  signal to end-nodes.
The end nodes extract the desired packets by using the signal forwarded by the relay, information on the packet previously transmitted by themselves, and channel state information (CSI) of the relayed links. PLNC appears in several flavors, depending on the operation done at the relay, such as Amplify-and-Forward (AF)~\cite{PLNC_pp_hy}, Denoise-and-Forward (DNF) \cite{PopovskiICC}, etc.   

While many works on PLNC have been successfully showing its gain for two-way relaying in an isolated three-node model, some of the recent studies have attempted to employ PLNC in wireless networks of a larger scale\cite{Yoshiki}\cite{Wang}.
In such networks, there are many neighboring nodes surrounding the three nodes involved in PLNC, which may attempt to simultaneously access the shared channel and thus cause interference. 
A standard way to alleviate the impact of interference is to protect areas around transmitter/receiver from interfering nodes by control frames like RTS/CTS defined in IEEE802.11\cite{11MAC}\cite{Harald}.
On one hand, PLNC needs less time to exchange data frames than the one-way conventional relaying (CR)\cite{PLNC_pp_hy}\cite{PopovskiICC}.
On the other hand, in PLNC, the data exchange involves transmission/reception by three nodes, which means that PLNC needs a larger reserved area than CR, where only two nodes are involved in each step of data exchange.
Such a large reserved area diminishes the spatial reuse, which should be accounted for when evaluating PLNC in a large-scale wireless network.

In this paper, we analyze theoretically the end-to-end rate of PLNC and CR considering interference and spatial cost caused by the reserved area.
We derive the statistics of the interference caused by neighboring nodes that are outside the reserved areas. The numerical results highlight the impact of interference density and size of reserve area on the achievable rate, and reveal that the gains of PLNC over CR are not overwhelming and even may vanish when taking the spatial reuse into account.

\section{System Model and Problem Formulation}
\subsection{Transmission Modes\label{2.1}}

\begin{figure}
 \centering
 \includegraphics[width=5cm,height=3.5cm]{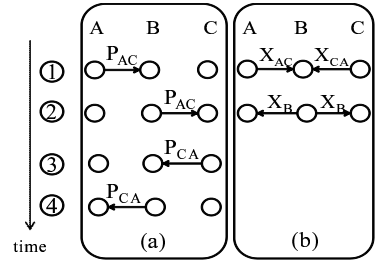}
 \caption{Packet Exchange (a) Conventional Relaying (b) PLNC}
 \label{fig1}
\end{figure}

Fig. \ref{fig1} illustrates two transmission modes considered in this paper: CR and PLNC.
Fig. \ref{fig1}(a) shows CR where nodes A and C exchange packets through relay node B.
Here, the packets $P_{AC}$ and $P_{CA}$ are destined from A to C and C to A, respectively.
As shown in Fig. \ref{fig1}(a), CR needs 4 time slots to exchange these packets.
On the other hand, Fig. \ref{fig1}(b) shows PLNC.
Let $x_{AC}$ and $x_{CA}$ be the complex baseband representations of the packets $P_{AC}$ and $P_{CA}$, respectively.
In the first slot, both A and C transmit signals simultaneously and the signals are added at the receiver through the multiple access channel. 
In this paper, we focus on PLNC with Amplify-and-Forward (AF), such that B amplifies the received signal and broadcasts $x_{B}$,  in the second slot\cite{PLNC_pp_hy}\cite{analog}.
Then, A and C attempt to decode their desired signals by eliminating self-interference component (i.e., $x_{CA}$ = $x_{B}$ -  $x_{AC}$).
Thus, PLNC requires half of the time to complete packet exchange as compared with CR in an ideal (noiseless) case.

\subsection{Node Deployment and Reserved Area\label{2.2}}

\begin{figure}
 \centering
 \includegraphics[width=6cm,height=5cm]{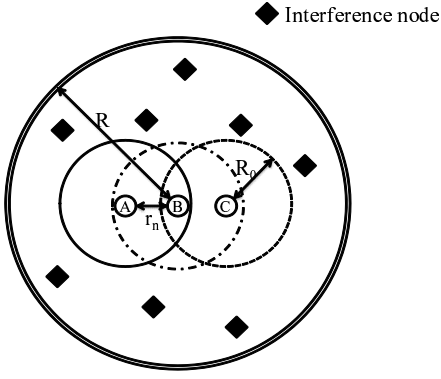}
 \caption{Node Deployment}
 \label{fig2}
\end{figure}

We apply PLNC to a large-scale wireless network, where many neighboring nodes simultaneously access the shared channel.
Fig. \ref{fig2} shows a network model used for the analysis in this paper.
We assume two-way relaying between A and C through relay node B.
However, these three nodes are surrounded by interfering nodes which attempt to transmit in the same channel.
A, B and C are positioned on a straight line, equally separated with distance $r_{n}$.
The most common method to control interference from neighboring nodes is channel reservation using control frames.
We assume that the nodes A, B and C send RTS/CTS-like control frames before data transmission to protect the area around them from interfering nodes.
For instance, the circle shown by the solid line in Fig. \ref{fig2} denotes the area reserved by the control frame transmitted by A.
Analogously, the circles shown by dashed and dotted lines denote the area reserved by control frames transmitted by B and C, respectively.
We assume that the transmission range of control frames are all equal, expressed by the radii of three circles $R_{0}$ in Fig. \ref{fig2}.
Interfering nodes are distributed inside a ring with an outer radius $R$, which is shown by doublet in Fig. \ref{fig2}, but outside the reserved areas.
We assume that the interfering nodes are located randomly and uniformly with  density $\lambda$ [nodes per unit area]\cite{Natasha}.
Furthermore, interfering nodes are fully-loaded and always transmitting interference signals.

\subsection{Channel Model and Maximum Rate\label{2.3}}
In this work, we employ path-loss only model as a model of signal propagation, with a path-loss coefficient of $4$\cite{Goldsmith}.
We define the normalized distance of 1 as a distance achieving the link SNR (Signal-to-Noise Ratio) of 0 dB.
Then, with the assumption on the same transmit and noise power at all the nodes, SNR of the arbitrary link can be obtained by scaling based on the distance between nodes.
The SINR (Signal-to-Interference plus Noise Ratio) at a receiver is expressed as
\begin{equation}
SINR = \frac{P_{R}}{N+I} = \frac{SNR}{1+INR} \: , \label{SINR}
\end{equation}
where $P_{R}$ is the received power, $N$ is the noise power, $I$ is the interference power, and INR is Interference-to-Noise Ratio.
INR at each interfering link can be also obtained based on the distance between an interfering node and a receiver by scaling from SNR at the normalized distance.
With SINR, the maximum rate, $R_{max}$, at which information can be transmitted per unit frequency is calculated as\cite{Shannon}
\begin{equation}
R_{max} = \log_2 (1+SINR)\ \mathrm{[bit/s/Hz]} \: . \label{Rate}
\end{equation}

\subsection{Problem Formulation\label{2.4}}
As the reserved area becomes larger, the adverse effect of interference nodes is reduced.
However, this reduces the reusability of radio resource since there are more nodes inside the reserved area that cannot transmit data frames.
This implies that there is a trade-off between an effect of control frame for reducing the interference from interfering nodes and the reusability of radio resource.
In order to take this trade-off, we evaluate the achievable rate per unit area which is calculated considering spatial cost caused by the reservation made by control frames transmitted by nodes A-C. 
Compared to CR, in PLNC, the reserved area must be larger due to multiple transmissions that take place in the two-way relaying.
Such a larger reservation area incurs a larger spatial cost compared with CR, and thus reducing the gain usually observed in PLNC.
Therefore, in the following sections, we evaluate and compare the achievable rate of PLNC and CR considering the spatial cost of the reserved area.

\section{Analysis of the End-to-End Rate\label{3}}
In this section, we theoretically analyze end-to-end rate of PLNC and CR.
We first derive statistics of interference from nodes located outside the reserved area to obtain INR required to calculate each link rate.
Then, we introduce end-to-end rate considering the spatial cost of reserved area.

\subsection{Interference Statistics\label{3.1}}

\begin{figure}
 \centering
 \includegraphics[width=8cm,height=4.7cm]{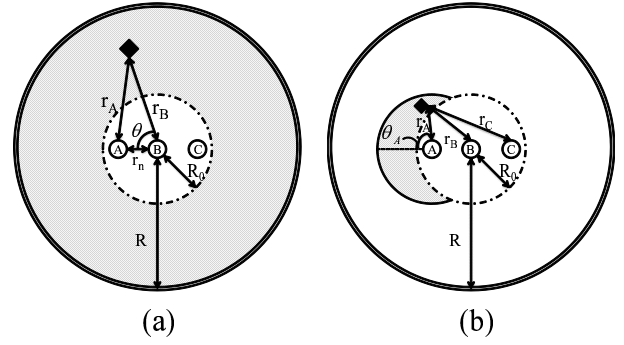}
 \caption{(a) Interference from Toroidal Area (b) Interference from Crescent Area}
 \label{fig3}
\end{figure}

In order to derive INR observed by different nodes with different shapes of the reserved area, we need to first consider interference from several areas inside the network ring.
We first evaluate the interference from the nodes located in the toroidal area to B, see Fig. \ref{fig3}(a).
We consider interfering nodes located uniformly within the toroidal area.
Then, the distance $r_{B}$ between an interfering node and B has the density
\begin{equation}
f_{r_{B}}(r_{B}) = \frac{2r_{B}}{R^{2} - R_{0}^{2}} \: , \:\: R_{0} \le r_{B} \le R \: . \label{r_B}
\end{equation}
INR observed at B due to interference from this node can be calculated as $\frac{1}{r_{B}^{4}}$ considering the normalized SNR of 0 dB at the distance 1.
The total number of interfering nodes within the toroidal area is $n^{toro}$ = $\lambda\pi$($R^{2}$ - $R_{0}^{2}$).
Therefore, the expected value of total INR at B is calculated as
\begin{eqnarray}
INR_{B}^{toro} &= n^{toro}\int_{R_{0}}^{R} \frac{f_{r_{B}}(r_{B})}{r_{B}^{4}} dr_{B} \nonumber\\
&= \pi\lambda \Bigl(\frac{1}{R_{0}^{2}} - \frac{1}{R^{2}} \Bigr) \: . \label{INR_B}
\end{eqnarray}
With $R$ $\to$ $\infty$, this INR becomes
\begin{equation}
INR_{B,\infty}^{toro} = \frac{\pi\lambda}{R_{0}^{2}} \: . \label{INR_B_Infty}
\end{equation}
To evaluate the interference from the toroidal area to A, we use the 
$r_{A}$($r_{B}$, $\theta$) between an interfering node and A:
\begin{equation}
r_{A}(r_{B}, \theta)^{2} = r_{B}^{2} + r_{n}^{2} - 2r_{n}r_{B}\cos\theta \: .  \label{d_r}
\end{equation}
Here, the joint probability density function, $f(r_{B}, \theta)$, is given by $\frac{r_{B}}{\pi(R^{2} - R_{0}^{2})}$.
The expected value of the total INR at A is calculated as
\begin{eqnarray}
INR_{A}^{toro} &= n^{toro}\int_{R_{0}}^{R} \int_{0}^{2\pi} \frac{f(r_{B}, \theta)}{r_{A}(r_{B}, \theta)^{4}} d\theta dr_{B} \nonumber\\
&= \pi\lambda \biggl\{\frac{R_{0}^{2}}{( R_{0}^{2} - r_{n}^{2})^{2}} - \frac{R^{2}}{(R^{2} - r_{n}^{2})^{2}} \biggl\} \: . \label{INR_A}
\end{eqnarray}
Next, we evaluate the interference from a crescent area to A as shown in Fig. \ref{fig3}(b).
$S_{cre}$ denotes the size of crescent area.
The number of interfering nodes within this area is $n^{cre} = \lambda$ $S_{cre}$.
With uniform distribution of nodes within the crescent area, the distance $r_{A}$ between an interfering node and A has the density
\begin{equation}
f_{r_{A}}(r_{A}) = \frac{2r_{A}\varphi(r_{A})}{S^{cre}} \: , \:\: R_{0} - r_{n} \le r_{A} \le R_{0} \: , \label{r_A}
\end{equation}
where $\varphi(r_{A})$ = $\cos^{-1}\bigl(\frac{R_{0}^{2}-r_{n}^{2}-r_{A}^{2}}{2r_{A}r_{n}}\bigr)$.
Therefore, the expected total INR at A is calculated as
\begin{equation}
INR_{A}^{cre} = n^{cre}\int_{R_{0}-r_{n}}^{R_{0}} \frac{f_{r_{A}}(r_{A})}{r_{A}^{4}} dr_{A} \: . \label{INR_A_can}
\end{equation}
Analogously, we evaluate the interference from the crescent area to B.
The distance $r_{B}$($r_{A}$, $\theta_{A}$) between this interfering node and B satisfies
\begin{equation}
r_{B}(r_{A}, \theta_{A})^{2} = r_{A}^{2} + r_{n}^{2} + 2r_{n}r_{A}\cos\theta_{A} \: .    \label{d_r_B}
\end{equation}
Here, the joint probability density function, $f(r_{A}, \theta_{A})$, is $\frac{r_{A}}{S_{cre}}$.
The expected total INR at B is calculated as
\begin{equation}
INR_{B}^{cre} = n^{cre}\int_{R_{0}-r_{n}}^{R_{0}} \int_{-\varphi(r_{A})}^{\varphi(r_{A})} \frac{f(r_{A}, \theta_{A})}{r_{B}(r_{A}, \theta_{A})^{4}} d\theta_{A}dr_{A} \: .  \label{INR_B_can}
\end{equation}
Finally, we analyze the interference from the crescent area to C.
The distance $r_{C}$($r_{A}$, $\theta_{A}$) between an interfering node and C satisfies
\begin{equation}
r_{C}(r_{A}, \theta_{A})^{2} = r_{A}^{2} + 4r_{n}^{2} + 4r_{n}r_{A}\cos\theta_{A} \: .    \label{d_r_C}
\end{equation}
The expected value of total INR at C is calculated as
\begin{equation}
INR_{C}^{cre} = n^{cre}\int_{R_{0}-r_{n}}^{R_{0}} \int_{-\varphi(r_{A})}^{\varphi(r_{A})} \frac{f(r_{A}, \theta_{A})}{r_{C}(r_{A}, \theta_{A})^{4}} d\theta_{A}dr_{A} \: . \label{INR_C_can}
\end{equation}

\subsection{INR for PLNC and CR\label{3.2}}

\begin{figure}
 \centering
 \includegraphics[width=8cm,height=5cm]{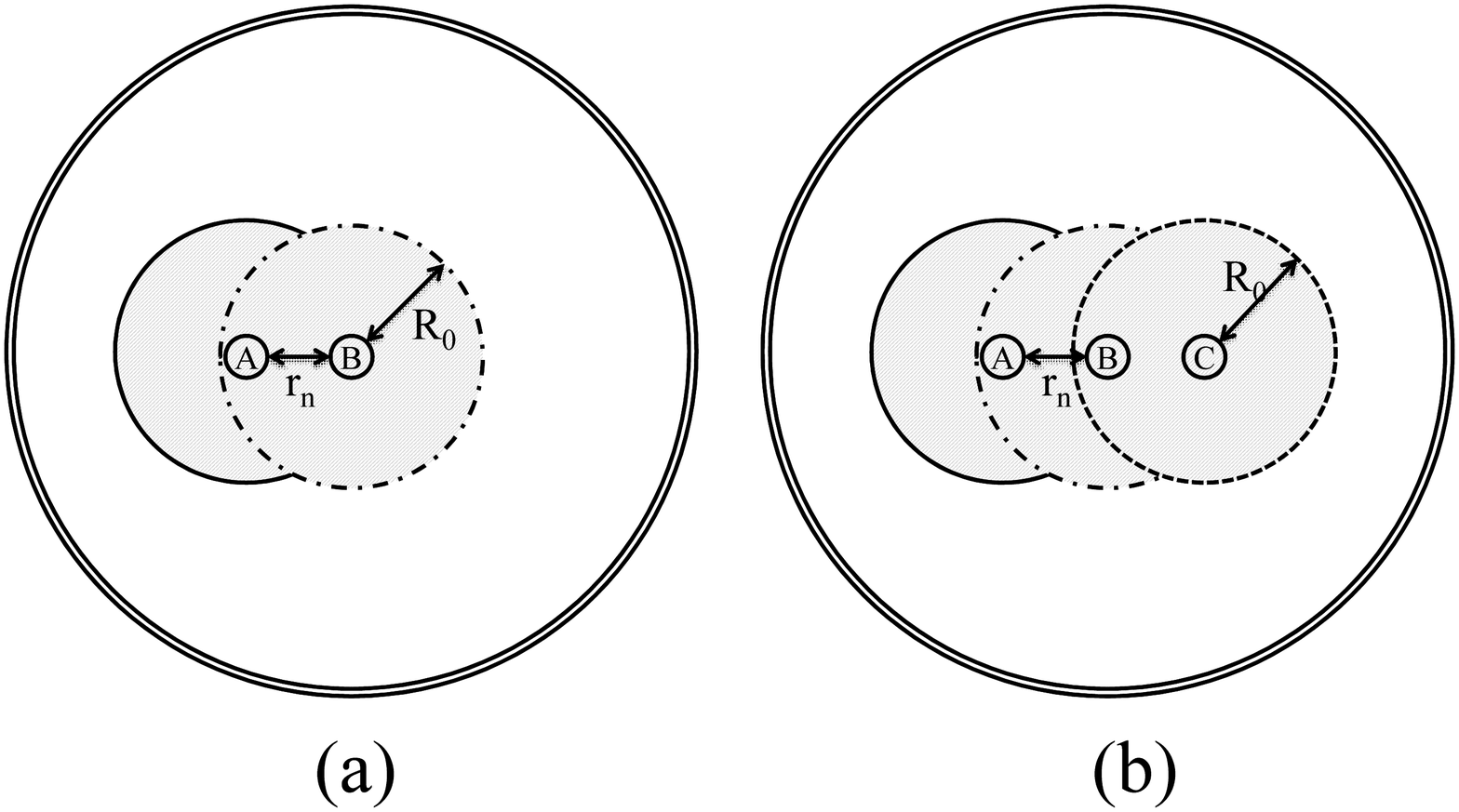}
 \caption{(a) Reserved area in the first slot for CR (b) Reserved area for PLNC}
 \label{fig4}
\end{figure}

Based on the interference statistics derived in the previous subsection, we derive INR for PLNC and CR when the transmitter and receiver have reserved area around them.
For CR, in the first slot, A transmits a packet to B, and the reserved area exists around these two nodes as shown in Fig. \ref{fig4}(a).
The total interference can be obtained by subtracting interference due to crescent area from that due to toroidal area and the expected total INR at B is expressed as
\begin{equation}
INR_{B}^{CR} = INR_{B}^{toro} - INR_{B}^{cre} \: ,  \label{INR_B_R}
\end{equation}
where INR$_{B}^{toro}$ and INR$_{B}^{cre}$ are given by (\ref{INR_B}) and (\ref{INR_B_can}), respectively.
Analogously, the expected INR at A is given by
\begin{equation}
INR_{A}^{CR} = INR_{A}^{toro} - INR_{A}^{cre} \: ,  \label{INR_A_R}
\end{equation}
where INR$_{A}^{toro}$ and INR$_{A}^{cre}$ are given by (\ref{INR_A}) and (\ref{INR_A_can}), respectively.
Due to the symmetry, all the INRs in the different time slot can be similarly obtained.
For instance, INR at node C in the second slot is same as (\ref{INR_A_R}), i.e., INR$_{A}^{CR}$ = INR$_{C}^{CR}$.
In the case of PLNC, all the three nodes are involved in transmission/reception in both slots, therefore, the reserved area should be configured as in Fig. \ref{fig4}(b).
In this case, the expected total INR at B is expressed as
\begin{equation}
INR_{B}^{PLNC} = INR_{B}^{toro} - 2 \times INR_{B}^{cre} \: . \label{INR_B_PLNC}
\end{equation}
Similarly, the expected total INR at A is given by
\begin{equation}
INR_{A}^{PLNC} = INR_{A}^{toro} - INR_{A}^{cre} - INR_{C}^{cre} \: . \label{INR_A_PLNC}
\end{equation}
where INR$_{C}^{cre}$ is given by (\ref{INR_C_can}).
Here, we used the fact that the INR at A from a crescent area in the right-hand side is the same as that at C from the crescent area in the left-hand side, i.e., INR$_{C}^{cre}$.
Due to the symmetry, INR$_{A}^{PLNC}$ = INR$_{C}^{PLNC}$.

\subsection{End-to-End Rate\label{3.3}}
In this section, we analyze end-to-end rate considering spatial cost due to reserved area.
Let $T_{slot}$ [symbols] be one slot time.
In this paper, we neglect the overhead in terms of time required to transmit control frames, and assume that the whole slot is used for data transmission. 

We first consider CR.
The SINRs at link B-A, A-B, B-C and C-B are denoted as $\gamma_{BA}$, $\gamma_{AB}$, $\gamma_{BC}$ and $\gamma_{CB}$, respectively.
These SINRs can be obtained from (\ref{SINR}) with given SNR and INR derived in the previous subsection.
Then, $R_{AB}$, $R_{BC}$, $R_{CB}$ and $R_{BA}$, which are the rate of link A-B, B-C, C-B and B-A, respectively, can be calculated with (\ref{Rate}).
The size of reserved area in each slot shown in Fig. \ref{fig4}(a), $S_{CR}$ is
\begin{equation}
S_{CR} = 2R_{0}^{2} (\pi - \psi) + \frac{r_{n}}{2}\sqrt{4R_{0}^{2} - r_{n}^{2}} \: , \label{S_R}
\end{equation}
where $\psi = \tan^{-1}\Bigl(\frac{\sqrt{4R_{0}^{2} - r_{n}^{2}}}{r_{n}}\Bigr)$.
We consider that this area is the spatial cost in each time slot, therefore, total temporal/spatial cost in each slot is $S_{CR}T_{slot}$.
Then, end-to-end rate per unit time/area in the bidirectional communication with CR can be expressed as
\begin{eqnarray}
R^{CR} &= \frac{\min (R_{AB}T_{slot} , R_{BC}T_{slot}) + \min (R_{CB}T_{slot} , R_{BA}T_{slot})}{4S_{CR}T_{slot}} \nonumber\\
 &= \frac{\min (R_{AB} , R_{BC}) + \min (R_{CB} , R_{BA})}{4S_{CR}} \: . \label{shannon_r}
\end{eqnarray}
Subsequently, we consider PLNC.
SINRs observed at destination nodes, A and C, when AF PLNC is applied, depend on SINRs at each link, and can be respectively expressed with $\gamma_{BA}$, $\gamma_{AB}$, $\gamma_{BC}$ and $\gamma_{CB}$ as
\begin{equation}
\gamma_{A} = \frac{\gamma_{BA}\gamma_{CB}}{1+\gamma_{BA}+\gamma_{AB}+\gamma_{CB}} \: , \label{rA}
\end{equation}
\begin{equation}
\gamma_{C} = \frac{\gamma_{AB}\gamma_{BC}}{1+\gamma_{AB}+\gamma_{BC}+\gamma_{CB}} \: . \label{rC}
\end{equation}
$R_{CA}^{PLNC}$ and $R_{AC}^{PLNC}$ respectively denote the rate observed at A and C, which is calculated with (\ref{Rate}), (\ref{rA}) and (\ref{rC}).
The size of reserved area with PLNC in each slot shown in Fig. \ref{fig4}(b), $S_{PLNC}$, is calculated as
\begin{equation}
S_{PLNC} = R_{0}^{2} (3\pi - 4\psi) + r_{n}\sqrt{4R_{0}^{2} - r_{n}^{2}} \: . \label{S_PLNC}
\end{equation}
Then, end-to-end rate per unit time/area in the bidirectional communication with PLNC is expressed as
\begin{eqnarray}
{\scriptstyle R^{PLNC} = \frac{ R_{AC}^{PLNC}T_{slot} + R_{CA}^{PLNC}T_{slot} }{2S_{PLNC}T_{slot}} = \frac{ R_{AC}^{PLNC} + R_{CA}^{PLNC}}{2S_{PLNC}}} \: . \label{shannon_plnc}
\end{eqnarray}

\section{Numerical Results}
In this section, we present numerical results on end-to-end rate obtained by theoretical analysis made in the previous section and by computer simulation.

\subsection{Validity of Network Radius\label{4.1}}

\begin{figure}
 \centering
 \includegraphics[width=8.5cm,height=4cm]{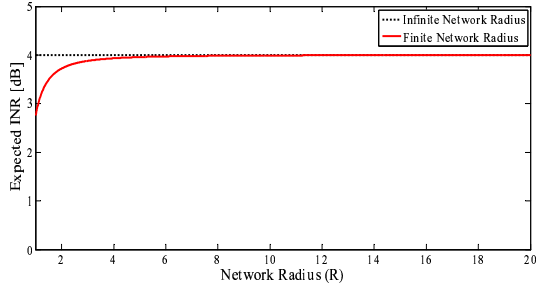}
 \caption{Expected INR with finite and infinite network radius for $R_{0}$ = 0.5, $\lambda$ = 0.2}
 \label{fig5}
\end{figure} 
In our system model, we employ a finite radius $R$ to determine the size of network within which all interfering nodes exist.
This radius should be large enough to include all dominant interfering nodes around nodes, A, B, and C.
Here, we first confirm the validity of network radius by comparing (\ref{INR_B}) with (\ref{INR_B_Infty}).
Fig. \ref{fig5} shows the expected total INR at B against the network radius when the radius of reserved area, $R_{0}=0.5$, with the interference density, $\lambda= 0.2$.
In Fig. \ref{fig5}, the dashed line indicates the results with infinite network radius, and the solid line with finite network radius, $R$.
From this figure, it can be seen that, as $R$ becomes larger, the corresponding INR approaches to the value with infinite radius, and the $R=10$ is large enough to include all the influence from interfering nodes.
Therefore, in this work, we fix the network radius $R=10$ further on. 

\subsection{Impact of the Reserved Area\label{4.2}}

\begin{figure}
 \centering
 \includegraphics[width=8.5cm,height=4cm]{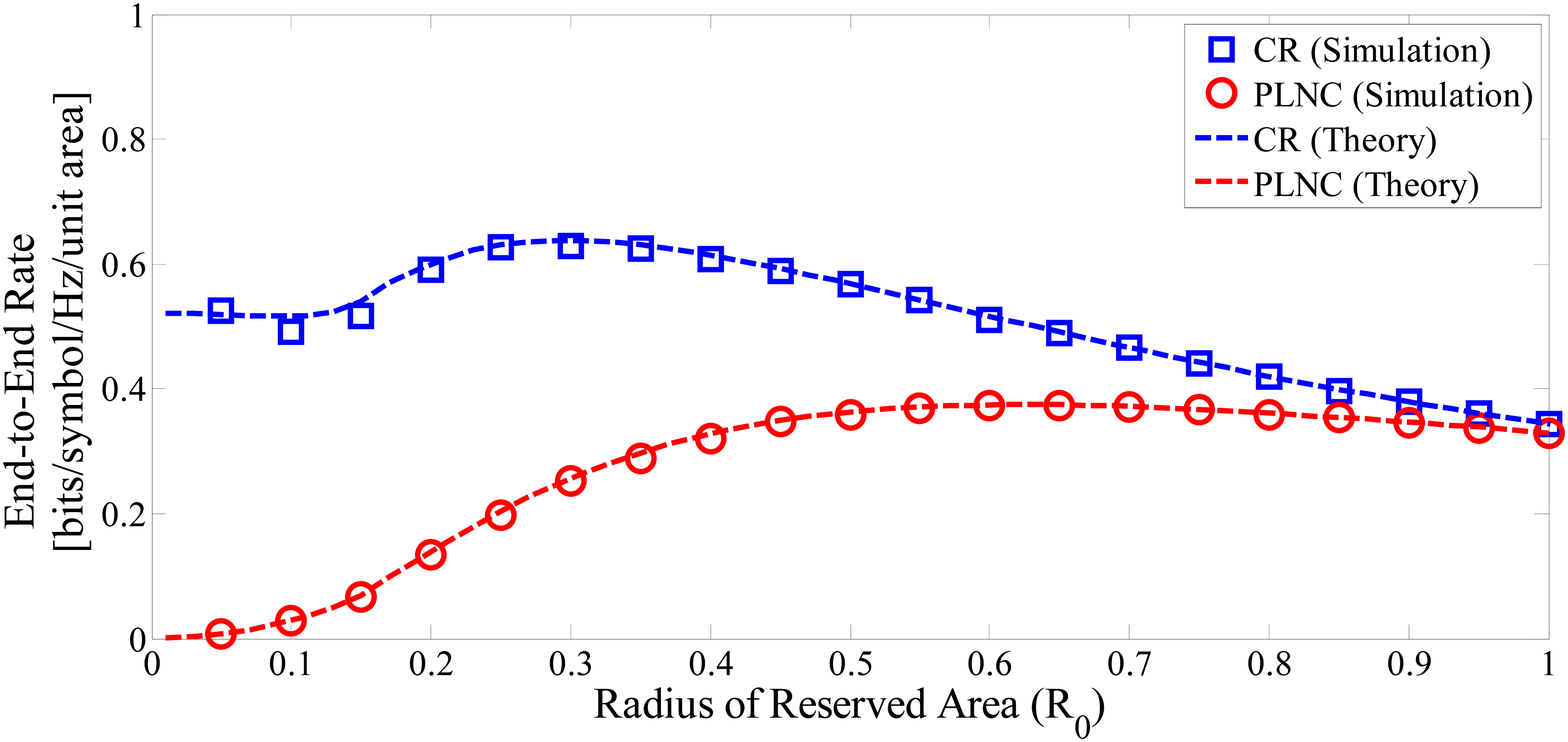}
 \caption{End-to-End Rate for link SNR = 20 dB, $\lambda$ = 7}
 \label{fig6}
\end{figure}
\begin{figure}
 \centering
 \includegraphics[width=8.5cm,height=4cm]{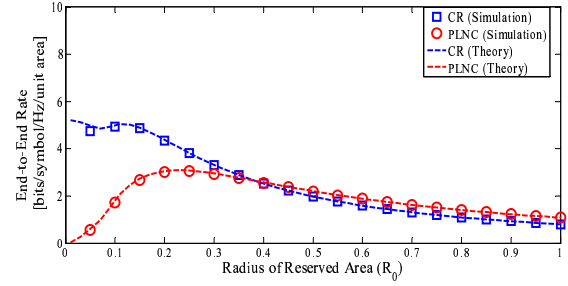}
 \caption{End-to-End Rate for link SNR = 30 dB, $\lambda$ = 7}
 \label{fig7}
\end{figure}

Fig. \ref{fig6} and Fig. \ref{fig7} show end-to-end rate of CR and PLNC against the radius of reserved area, $R_{0}$, with $\lambda=7$ [nodes / unit area].
Here, Fig. \ref{fig6} shows results when the link SNR (SNR at link A-B or B-C) is set to be 20 dB while Fig. \ref{fig7} shows results with 30 dB.
In these figures, lines indicate results calculated with equations obtained by theoretical derivations in previous section, and dots represent computer simulation results.
In general, the radius of reserved area should be larger than the distance between the corresponding transmitter and receiver, i.e., $R_{0}$ should be larger than the distances A-B and B-C.
The distance between neighboring nodes is 0.32 when the SNR at link A-B or B-C is set to be 20 dB (Fig. \ref{fig6}), 
and 0.18 for SNR with 30 dB (Fig. \ref{fig7}).
Hence, the radius of reserved area must be larger than 0.32 and 0.18 in Fig. \ref{fig6} and Fig. \ref{fig7}, respectively, which we call as the minimum radius of reserved area.
 
The analytical results are closely matching with the simulation, thus validating our theoretical derivation presented in the previous section.
This can validate our theoretical derivation presented in the previous section.
Next, we can see a trade-off between an effect to reduce interference and a spatial cost due to reservation for $R_{0}$ which is larger than the minimum radius of reserved area from Fig. \ref{fig6}.
For small reserved area, the spatial cost due to reservation is low, while the distance to the nodes causing interference is small.
As the reserved area grows, the interference impact decreases, but the spatial cost becomes too large to keep high end-to-end rate.
This suggests that there is an optimal size of reserved area in terms of end-to-end rate.
Furthermore, Fig. \ref{fig6} shows that CR has better performance for all sizes of reserved area since the relay attempts to decode the received data,  alleviating the negative impact of low link SNR and interference.
Moreover, CR has a high spatial efficiency as it reserves smaller area than PLNC, leading to a better end-to-end rate, despite the four slots required to exchange packets when link SNR is small as shown in Fig. \ref{fig6}.
On the other hand, Fig. \ref{fig7} shows that, when the SNR at each link is larger, CR has better performance only for smaller reserved area.
In PLNC, the relay node amplifies received signal which also contains the interference signals.
This deteriorates the SINR observed at destination nodes, leading to worse performance for smaller reserved area, where the impact of the interference is larger.
For a larger reserved area, the influence of interfering nodes is reduced, and the inherent gain PLNC in terms of less time slots wins over the adverse impact of interference and the large size of the reserved area.

Finally, we focus on the peak rate achieved by each scheme for the range larger than the minimum radius of reserved area in Fig. \ref{fig6} and Fig. \ref{fig7}.
We can see that the value of $R_{0}$ which gives the best end-to-end rate in PLNC is larger than CR.
Moreover, for larger link SNR (Fig. \ref{fig7}), the optimum $R_{0}$ is smaller than the case for weaker link SNR (Fig. \ref{fig6}), and approaches to the minimum radius of reserved area.
\subsection{Impact of Interference Density\label{4.3}}
\begin{figure}[t]
 \centering
 \includegraphics[width=8.5cm,height=4cm]{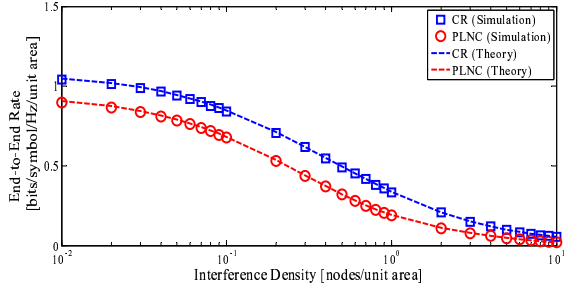}
 \caption{End-to-End Rate with the optimized size of reserved area for link SNR = 10 dB}
 \label{fig8}
\end{figure}
\begin{figure}[t]
 \centering
 \includegraphics[width=8.5cm,height=4cm]{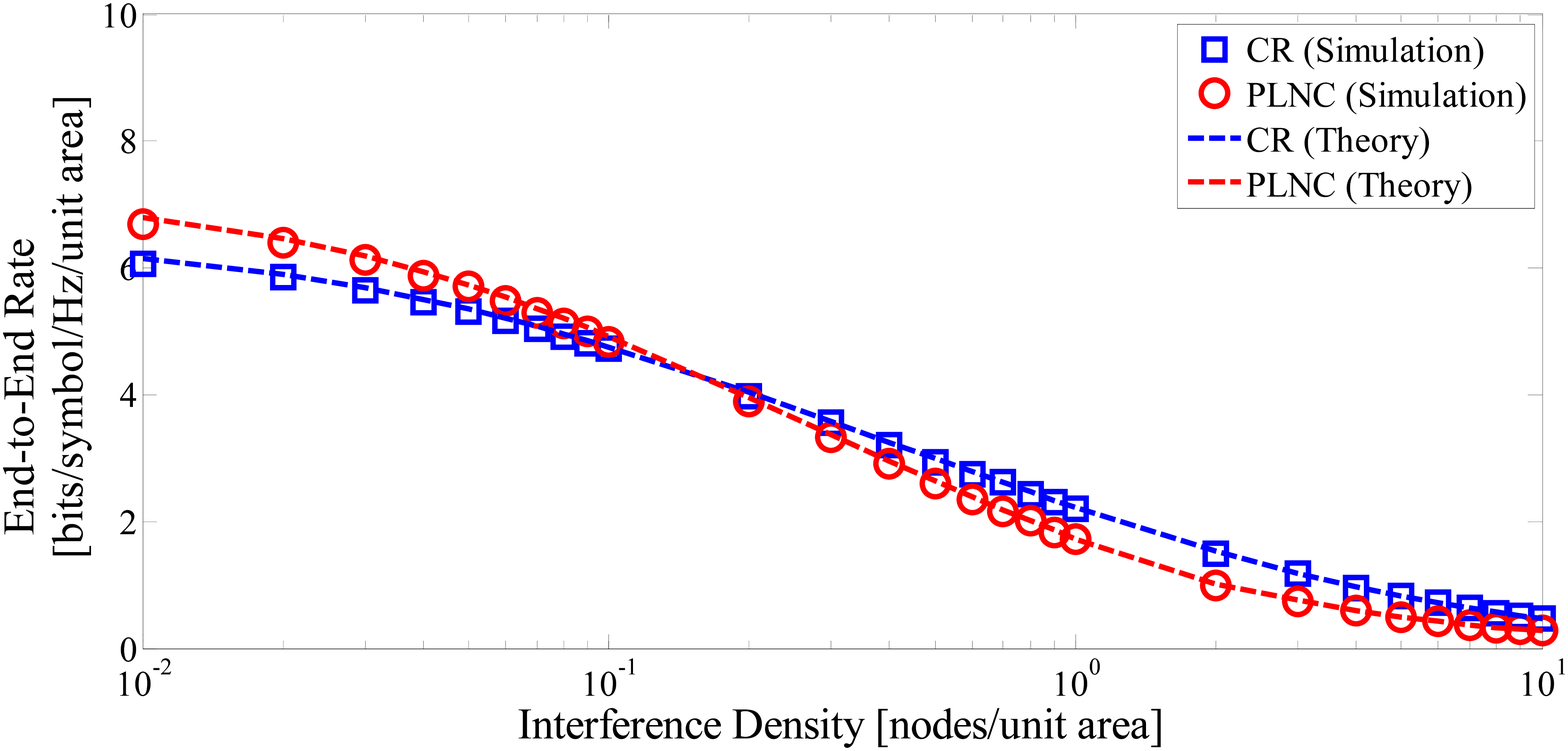}
 \caption{End-to-End Rate with the optimized size of reserved area for link SNR = 20 dB}
 \label{fig9}
\end{figure}
\begin{figure}[t]
 \centering
 \includegraphics[width=8.5cm,height=4cm]{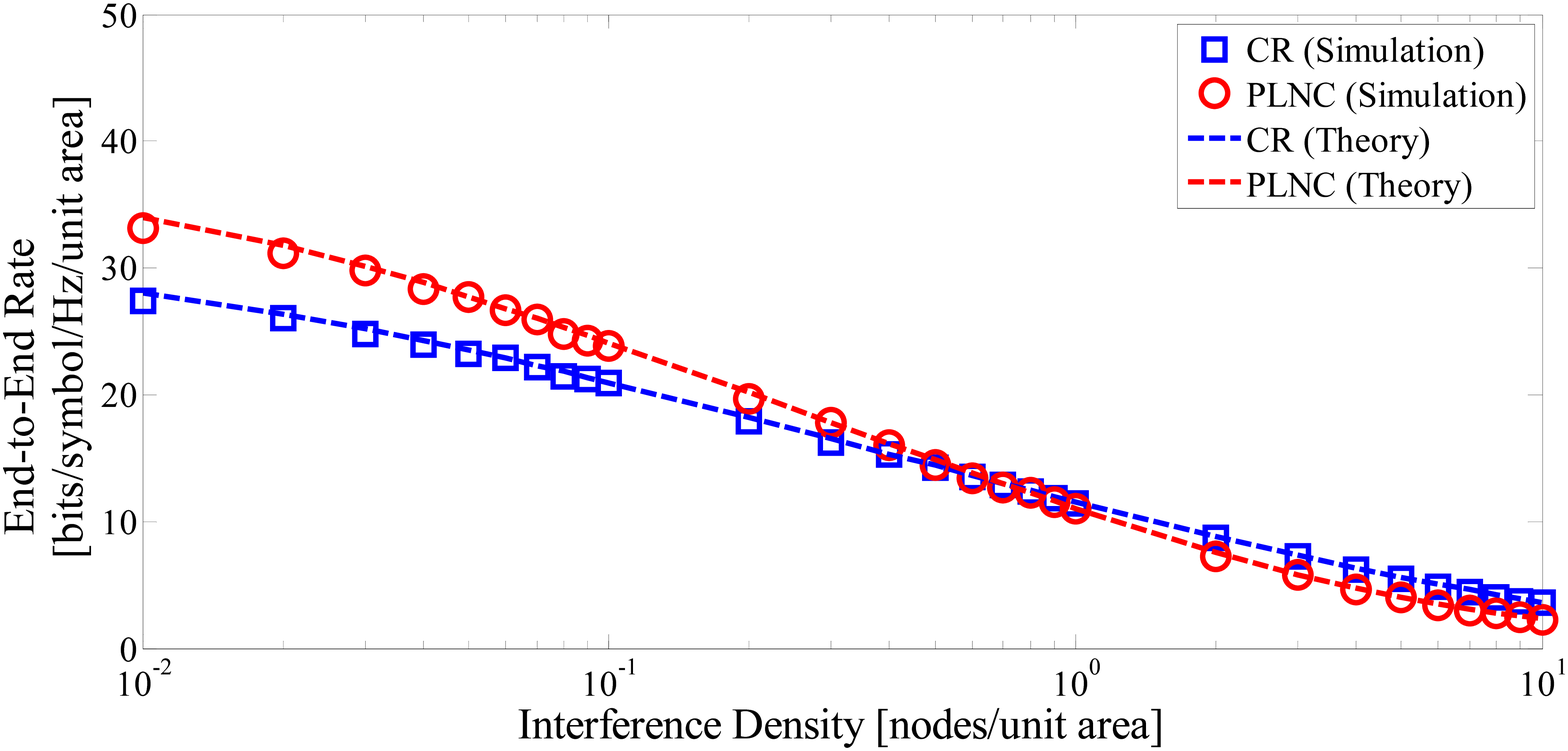}
 \caption{End-to-End Rate with the optimized size of reserved area for link SNR = 30 dB}
 \label{fig10}
\end{figure}
\begin{figure}[t]
 \centering
 \includegraphics[width=8.5cm,height=4cm]{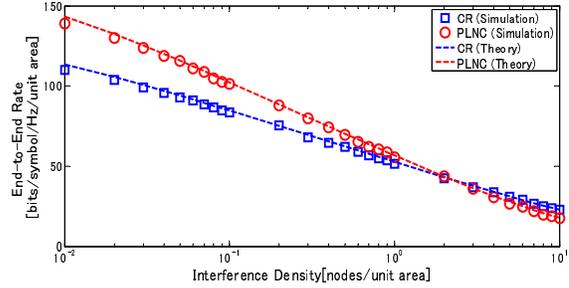}
 \caption{End-to-End Rate with the optimized size of reserved area for link SNR = 40 dB}
 \label{fig11}
\end{figure}
Next, we pay attention to end-to-end rate with the optimized size of reserved area for different link SNR and interfering node density.
Figs. \ref{fig8}-\ref{fig11} show the best achievable end-to-end rate of PLNC and CR against the interference density when the link SNRs are set to be 10, 20, 30, and 40 dB, respectively.
For each link SNR and interference density, we first calculate end-to-end rate against the size of reserved area, and obtain the best rate within the radius range larger than the minimum radius of reserved area.

It can be observed from Fig. \ref{fig8} that, when the SNR at neighboring nodes is small, CR outperforms PLNC in all the region.
This is because SINR at destination nodes in PLNC is low due to small link SNR and interference.
Furthermore, in this case, the optimum reserved area becomes larger for PLNC, which makes PLNC inferior to CR.
On the other hand, in Fig. \ref{fig9} with link SNR of 20 dB, we can see that PLNC outperforms CR for smaller interference density.
In fact, when interference density approaches to 0, PLNC shows more gain, which has been shown in many literatures investigating performance of PLNC in interference-free condition.
However, interestingly, CR shows better performance than PLNC as the interference density is increased.
This is due to interference lowering SINRs at destination nodes and larger spatial cost in PLNC.
This result clearly demonstrates the importance of analysis of PLNC considering the impact of interference.
Finally, Fig. \ref{fig10} and \ref{fig11} show that, as link SNR becomes larger, region where PLNC outperforms CR becomes larger.
In these cases, link SNR is so large that PLNC becomes more insensitive to interference.
These results clearly show the superiority of PLNC: when the link SNR is large enough, even with larger spatial cost, PLNC can outperform CR for a large range of interference density.

\section{Conclusions}
In this paper, we have analyzed end-to-end rate achieved by PLNC and CR considering the impact of interference and spatial cost due to reserved area in a large-scale wireless network.
First, we have shown a trade-off between the effect to reduce interference and reusability of radio resource, which is observed by changing the size of reserved area.
For large link SNR, the optimum radius of reserved area has been shown to be smaller than the case for small link SNR.
Then, we have analyzed the impact of interference density with the optimized size of reserved area.
We have shown that the superiority of PLNC is maintained for large SNR even with adverse effect of interference and larger reserved area than CR.

Our future work includes investigations with more general network model for PLNC nodes, e.g., unequal distance between A-B and B-C links.
It would be also interesting to employ different types of PLNCs, e.g., Decode-and-Forward and Denoise-and-Forward schemes\cite{PopovskiICC}.

\section*{Acknowledgment}
This work was in part supported by the JSPS Invitation Fellowship for Research in Japan (Short-Term) and by the Grant-in-Aid for Scientific Research (A) No. 24240009.

\end{document}